\documentclass[twocolumn,prl,aps,showpacs,floatfix]{revtex4}
\usepackage{graphicx,epsf}

\newcommand{\lsim}{\raise.3ex\hbox{$<$\kern-.75em\lower1ex\hbox{$\sim$}}}
\newcommand{\gsim}{\raise.3ex\hbox{$>$\kern-.75em\lower1ex\hbox{$\sim$}}}
\newcommand{\calC}{{\cal C}}

\begin{document}

\title{ A New Monte Carlo Method and Its Implications for Generalized 
Cluster Algorithms }

\author{C. H. Mak and Arun K. Sharma}

\affiliation{
  Department of Chemistry, 
  University of Southern California, \\
  Los Angeles, California 90089-0482, USA }

\date{\today}

\begin{abstract} 

We describe a novel switching algorithm based on a ``reverse'' Monte Carlo 
method, in which the 
potential is stochastically modified before the system configuration 
is moved.  This new algorithm facilitates a generalized 
formulation of cluster-type Monte Carlo methods, and the generalization 
makes it possible to derive cluster algorithms for systems with both 
discrete and continuous degrees of freedom.
The roughening transition in the sine-Gordon model has been 
studied with this method, and high-accuracy simulations for 
system sizes up to $1024^2$ were carried out to examine the 
logarithmic divergence of the surface roughness above the transition 
temperature, revealing clear evidence for universal scaling 
of the Kosterlitz-Thouless type.

\end{abstract}

\maketitle

Large-scale Monte Carlo (MC) simulations are often plagued by slow sampling 
problems.  These problems are especially severe in systems near the critical 
point or in those with strong correlations.  Slow sampling problems 
manifest themselves as 
poor scaling of the dynamic relaxation time with the system size, 
making large-size simulations extremely slow to converge.  The cause of 
these problems is that most MC simulations are based on local moves, and 
when the correlation length of the system grows or as relaxation modes 
of the system become heavily entangled, local moves become 
increasingly inefficient.  
But if nonlocal MC moves are used
\cite{ref1}, 
their acceptance ratios are often found to be 
exceedingly low when system correlations are strong.  

One way to circumvent these problems was suggested by 
Swendsen and Wang \cite{87swe86}, who 
devised a clever scheme where large-scale nonlocal MC moves may be 
constructed to achieve high sampling efficiencies by exploiting 
certain geometric symmetries in the system.  
This algorithm led to a marked reduction in the dynamical scaling exponent 
in the 2-dimensional Ising model near criticality.  
Since the nonlocal moves in this algorithm 
update a large number of degrees of freedom 
at the same time, the Swendsen-Wang method and others inspired by 
it are also often referred to as ``cluster Monte Carlo'' methods.

Since Swendsen and Wang's paper in 
1987, many cluster-type MC algorithms have appeared 
\cite{88edw2009, 88kan1591, 88wol1461, 88nie2026, 89wol361, 90kan941, 90wan565, 91kan8539, 91eve185, 92lia2145, 92has10472, 94has255, 95dre597, 01hou479, 02blo58, 03kra064503, 04liu035504}.  
But the success of cluster MC has not been universal because the proper 
cluster moves needed seem to be highly dependent on the system, and 
efficient cluster MC methods 
have been found for only a small number of models 
\cite{88edw2009,88wol1461,88nie2026,89wol361,91eve185,94has255,95dre597,03kra064503,04liu035504} 
so far.
The difficulty of formulating a generalized MC method that could work for 
any system seems to be associated with the apparent 
geometric nature of the cluster-type MC methods -- 
all existing cluster MC methods have been derived in one way or another 
by using certain 
geometric features of the system.  For example, in the original Swendsen-Wang 
formulation a mapping between the Ising model and the percolation 
model originally described by Fortuin and Kasteleyn \cite{72for536}
was exploited to effect cluster spin flips.  In other models, the 
requisite mapping is not always obvious, making cluster 
MC methods difficult to implement for general systems.

In this letter, we will show that the derivations of cluster MC methods 
do not have to be based on geometric features of the systems.  Instead,  
they may be more conveniently formulated based on algebraic features of the 
system potential $V(\calC)$.  
We will exploit this algebraic formulation and suggest a way
to generalize cluster Monte Carlo methods to systems with 
any potential.  

We focus on classical systems 
with partition function $Z = {\rm Tr}~e^{-V(\calC)}$, where 
$V(\calC)$ is the potential in units of the temperature $T$.
Acceptable Monte Carlo methods to sample the system configurations  
can be constructed using any transition probabilities 
$W(\calC \rightarrow \calC')$ 
as long as the detailed balance condition 
\begin{equation}
W(\calC \rightarrow \calC') e^{-V(\calC)} = W(\calC' \rightarrow 
\calC) e^{-V(\calC')}
\end{equation}
is satisfied.
Conventional MC methods such as Metropolis \cite{53met1087} accomplishes this 
in two steps: a trial move is made from $\calC$ to $\calC'$ with some 
transition probability, and the move is then accepted or rejected according to 
an acceptance probability based on $V(\calC')$, $V(\calC)$ or 
both, so that the composite process satisfies Eqn.(1).
This way of constructing the Markov chain -- trial moves followed by 
acceptance/rejection -- has been the accepted ``standard'' method 
for doing MC since the inception of the MC method \cite{49met335}.  
Other MC methods do exist, such as the heat bath algorithm \cite{86bina}, 
which follow alternative strategies, but by far the standard method is 
conceptually the simplest and most convenient in practice.

In the Monte Carlo method we are proposing, we will {\em reverse 
the order of the two steps in the standard method}.  
That is, we will first determine an acceptable way to modify the potential 
$V$ and then find a transition $\calC \rightarrow \calC'$ 
that is consistent with the new potential.  
To our knowledge, the basic elements of this ``reverse MC'' idea 
were first suggested by Kandel {\em et al.} \cite{88kan1591}, 
who used it to stochastically remove 
interaction terms from the system's potential in an Ising model 
to arrive at an alternative derivation of the Swendsen-Wang method.  
The formulation of Kandel {\em et al.} was limited to discrete systems 
like the Ising model.
In the following, 
we will show how the reverse MC idea may be formulated more 
generally for any system, discrete or continuous, and how it may then 
be used as a framework to construct generalized cluster algorithms.

Consider a system with potential $V = \sum_i v_i + V_r$, 
consisting of a number of ``interaction terms'' 
$v_i$ plus a ``residual'' $V_r$.
These interactions may be bonds between particles, interactions of 
the particles with a field, or any other additive terms in $V$.  
We consider replacing each interaction term 
$v_i$ by some {\em pre-selected} $\tilde v_i$ with a ``switching'' 
probability 
\begin{equation}
S_i(\calC) = c_i e^{\Delta v_i(\calC)},
\end{equation}
where $\Delta v_i = v_i - \tilde v_i$, $c_i = e^{-\Delta v_i^*}$ and 
$\Delta v_i^* = \max_{\calC}~v_i(\calC)$.
The outcome of the switches defines two complementary sets of interactions
-- the 
switched ones $\tilde \sigma$ and the unswitched ones $\bar \sigma$.  
Using the outcome of the switches, we define a stochastically 
modified potential $\tilde V$ as follows:
\begin{equation}
\tilde V = \sum_{i\in \tilde \sigma} \tilde v_i 
+ \sum_{j\in \bar \sigma} \bar v_j + V_r, 
\end{equation}
with $\bar v_i = v_i - \ln (1-S_i)$.  
An MC pass starts with an attempt to switch every interaction $v_i$
to the new $\tilde v_i$ using the 
$S_i$ defined in Eqn.(2).  
If the switch is successful, the interaction is replaced by $\tilde v_i$.  
If not, the interaction is replaced by another interaction $\bar v_i$.  
This is followed by an update in the 
configuration of the entire system 
from $\calC$ to $\calC'$ using a transition probability 
$\tilde W(\calC \rightarrow \calC')$ that satisfies detailed balance 
{\em on the modified potential} $\tilde V$.  
This constitutes one pass.  The move from $\calC$ to $\calC'$ can 
of course be carried out using any conventional MC move that satisfies 
detailed balance on the modified potential.  But the reverse 
formulation of the MC method now offers possibilities that were 
previously unavailable to conventional MC methods --- if a simple 
scheme can be devised to update the configuration of the {\em entire} 
system on the stochastically modified potential, one can envision 
designing global moves for the system to accelerate its sampling, and 
our freedom in choosing the $\tilde v_i$ can be actively exploited 
to facilitate this.  
Within this context, the original formulation of Kandel {\em et al.}
corresponds to switching $v_i$ to $\tilde v_i = 0$, i.e. 
simplifying the potential by deleting interactions from it.  
Kandel {\em et al.} showed that for the Ising model they could easily 
construct global moves on this stochastically simplified potential and 
their formulation regenerates the Swendsen-Wang method.  
But compared to the deletion formulation of Kandel {\em et al.}, 
the switching implementation of the reverse MC method now offers a much 
wider set of possibilities because the 
form of the ``switch to'' interactions is completely arbitrary.  
Whereas previously there may not be an obvious way to globally update 
the configuration of the system on the original potential, 
with the proper choices for $\tilde v_i$ 
large-scale moves may now become possible on the stochastically 
modified potential.  
Indeed, we have shown that the switching idea may be 
used to formulate a cluster MC algorithm for a Lennard-Jones 
fluid \cite{05mak214110}.  

Equations~(2), (3) and the transition probability $\tilde W$ define the 
switching algorithm.  
To prove detailed balance Eqn.(1) for the switching algorithm, it is 
sufficient to treat a case where there are only two interaction
terms.  Extension to any number of interactions is straightforward.
Starting with $\calC$, 
with two interaction terms $v_1$ and $v_2$, 
there are four possible outcomes from the switch: 
I. both 1 and 2 are switched, which occurs with probability 
$P_{\rm I} = S_1(\calC) S_2(\calC)$, 
II. 1 is switched and 2 is unswitched, with 
$P_{\rm II} = S_1(\calC) [1-S_2(\calC)]$, 
III. 1 is unswitched and 2 is switched, with 
$P_{\rm III} = [1-S_1(\calC)] S_2(\calC)$, and 
IV. both 1 and 2 are unswitched, with 
$P_{\rm IV} = [1-S_1(\calC)] [1-S_2(\calC)]$.  
After the switch, an update $\calC \to \calC'$ is made with a transition 
probability $\tilde W$ that satisfies detailed balance on the modified 
potential $\tilde V$ defined in Eqn.(3).  
Each of the four channels will 
have a different $\tilde W$: $\tilde W_{\rm I}$, $\tilde W_{\rm II}$, etc.,  
and $W(\calC \to \calC')$ in Eqn.(1) is the sum 
$
P_{\rm I} \tilde W_{\rm I} + 
P_{\rm II} \tilde W_{\rm II} + 
P_{\rm III} \tilde W_{\rm III} + 
P_{\rm IV} \tilde W_{\rm IV} 
$
over all four channels.  
For the reverse transition, we start 
with $\calC'$ and consider switching $v_1(\calC') \to \tilde v_1(\calC')$ 
and $v_2(\calC') \to \tilde v_2(\calC')$.  Again there are 
four possible outcomes and we call these scenarios 
I$'$, II$'$, III$'$ and IV$'$ 
as for the forward transition.  
$W(\calC' \to \calC)$ in Eqn.(1) is again the sum 
$
P_{\rm I'} \tilde W_{\rm I'} + 
P_{\rm II'} \tilde W_{\rm II'} +
P_{\rm III'} \tilde W_{\rm III'} +
P_{\rm IV'} \tilde W_{\rm IV'}
$.
Using the choice of $S$ and $\tilde V$ in Eqs.(2) and (3), 
it is easy to show that detailed balance is obeyed 
{\em along each channel}, i.e. 
$P_{\rm I} \tilde W_{\rm I} = P_{\rm I'} \tilde W_{\rm I'}$, 
$P_{\rm II} \tilde W_{\rm II} = P_{\rm II'} \tilde W_{\rm II'}$, etc.
Of course, detailed balance only requires the {\em total} $W$ to satisfy 
Eqn.(1), and it is possible to choose alternate forms of $S$ and $\tilde V$ 
to do that, which may provide further flexibilities.

In the rest of this letter, we will illustrate the effectiveness of 
the switching implementation of the reverse MC method, and show how it can be 
used to easily derive a cluster MC method in a system with 
continuous degrees of freedom.  Previously, it has been extremely difficult to 
design cluster MC algorithms for systems with continuous degrees of freedom.  
The few that have been reported to date 
\cite{88edw2009,88wol1461,88nie2026,89wol361,91eve185,94has255,04liu035504}
were mainly based on embedding discrete degrees of freedom into continuous 
ones.  The only exception is the recent discovery of a geometric MC 
algorithm by Liu and Luitjen \cite{04liu035504} 
where they formulated a rejection-free MC method to sample the Lennard-Jones 
fluid at its critical point.  

The switching algorithm we have proposed 
makes the process of deriving cluster-type MC methods much 
more straightforward compared to those based on geometric features of 
the system.  
We will illustrate this using the sine-Gordon model, which 
can be used to study the roughening transition on 2-dimensional surfaces.  
The sine-Gordon (SG) model has the potential 
\begin{equation}
V_{\rm SG} = T^{-1} \left[ 
\frac{1}{2} \sum_{\langle i,j\rangle} |\phi_i - \phi_j|^2
- \sum_i \cos(\phi_i) \right],
\end{equation}
where $\phi_i$ are continuous variables on a 2-dimensional square lattice, 
the second sum is over all sites  
and the first sum is over all nearest-neighbor pairs.  
The SG model is often considered to be a coarse-grained version of 
the discrete Gaussian (DG) model with potential 
$V_{\rm DG} = T^{-1} \sum_{\langle i,j\rangle} |h_i - h_j|^2,$ 
where $h_i$ are integers.  The DG model can in turn be mapped directly onto 
the Coulomb gas model \cite{76chu4978}, and as a result, the SG model should 
belong in the same universality class as the Kosterlitz-Thouless (KT) 
transition \cite{73kos1181,74kos1046}.  

Roughening is expected to be a weak transition.  
The only easily discernible divergence is exhibited in a logarithmic 
dependence of the surface roughness 
$\sigma^2 = \langle |\phi_i - \langle\phi\rangle|^2 \rangle$ 
on the system 
size $L$ at the roughening temperature $T_R$.  Below $T_R$, $\sigma^2$ is 
expected to approach a finite value as $L\to\infty$. 
In addition to this, since 
the divergence is slow, large lattice sizes are needed to reach the 
scaling limit.  All of these features of the SG model make it hard 
to accurately study the roughening transition using MC 
simulations.  Previous simulations have been limited to small 
systems \cite{77swe5421,78shu1399,91fal8081,94has255,95san14664,00san3219}.  

In order to locate $T_R$ and study the scaling behavior at the 
roughening transition, we make use of the switching algorithm of 
the reverse MC method proposed above.  The essential difficulty in 
treating the SG model is due to the nonlinear cosine terms in the 
potential in Eqn.(4).  If these nonlinearities could be removed, the 
residual potential becomes a simple Gaussian and we could move 
the system configuration effectively using uncoupled surface modes.  
With this in mind, we separate the potential into two parts and treat 
the cosine terms as interactions $v_i = -T^{-1}\cos\phi_i$ and the 
harmonic part as the residual $V_r$.  Each of the interactions is switched to 
a uniform potential $\tilde v_i=-T^{-1}$ with $S_i = e^{[1-\cos\phi_i]/T}$.
After the switches, a number of $\phi_i$ would have effectively 
lost their couplings to the cosine potential, while the rest have 
their interactions with the cosine potential replaced by 
$\bar v_i = -\ln [ e^{\cos\phi_i/T} - e^{-1/T} ]$.  
In the ensuring MC move, we can update the unswitched $\phi_i$ which are now 
coupled to the replacement interactions $\bar v_i$ using conventional methods, 
but try to formulate an 
update scheme where the rest of the $\phi_i$, now forming a constrained 
Gaussian field, may be updated globally.
A Gaussian field subject to linear constraints is still 
Gaussian, and in principle 
we can diagonalize the potential to obtain all the normal 
modes and then move each one independently.  This problem 
is the subject of fracton dynamics and has been studied previously 
\cite{94nak381}.  
However, the cost of obtaining all the normal modes of the constrained 
surface and their frequencies will grow rapidly with the size of the lattice 
and will only be feasible for small-size simulations.  Since the scaling 
limit in the SG model can only be reached with large system sizes, we will 
need an alternative method.
The method we have used to update the constrained Gaussian fields 
is based on the method of Hoffman and Ribak \cite{91hof5}.  
Since the statistics of the fluctuations of a Gaussian field from its mean 
is independent of the value of the mean field, the fluctuations 
from a free Gaussian field can be transferred to a constrained field with 
a different mean.
Near the roughening temperature, the switching procedure produces 
roughly 5\% unswitched field points, and the corresponding mean field with these 
constraints can easily be determined using a steepest descent molecular 
dynamics method.  
To ensure ergodicity, a conventional Metropolis move is also carried out 
with every reverse MC move.

\begin{figure}
\includegraphics[width=1.0\columnwidth]{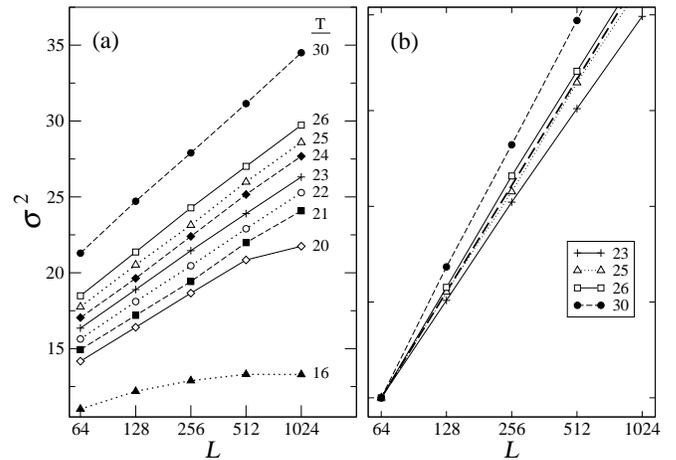}
\caption[]{
(a) Surface thickness $\sigma^2$ as a function of the log of 
lattice size $L$ for different temperatures $T$.  
(b) Expanded view of (a) for several temperatures near $T_R$ 
shifted vertically to coincide at $L=64$.  Dashed line is 
the expected KT slope at $T_R$, showing that $T_R$ is slightly 
above $T=25$ but below 26.
}
\end{figure}

Figure~1(a) shows simulation results for the scaling of the surface 
roughness $\sigma^2$ with the length $L$ of the lattice 
in simulations with different lattice sizes $L^2$ up to $1024^2$ and at 
several temperatures $T$ from 16 to 30.  
KT theory \cite{73kos1181,74kos1046} 
predicts a logarithmic divergence for $\sigma^2$ 
with a universal slope at $T_R$
\begin{equation}
\sigma^2(L) = \sigma_0^2(T_R) + \frac{a^2}{\pi^2} \ln L, 
\end{equation}
where $a$ is the lattice constant of the surface, and in the units
of Eqn.(4), $a = 2\pi$.  
Therefore, at $T_R$ the slope of Fig.~1(a) should be equal to 4.
Above $T_R$, the logarithmic behavior of $\sigma^2$ continues to hold 
except the constant $\sigma_0^2$ as well as the slope both 
increase with $T$.
The data in Fig.~1(a) show that for $T = 21$ and below, 
$\sigma^2$ appears to approach a finite value as $L \to \infty$.
Therefore, it is clear that $T_R > 21$.  The most recent simulation 
of the SG model by Sanchez {\em et al.} \cite{00san3219} (referred to 
as the ``ordered SG model'', OSGM, in this paper)  
suggested that $T_R \approx 16$.  Our data show that this is incorrect, 
and their error is likely due to slow sampling problems.
Locating the precise value of $T_R$ is more involved, since the 
data for $T > 21$ show no obvious tendency toward a finite $\sigma^2$.  
There are two possibilities: either these temperatures 
are above $T_R$ or the system size may not be 
large enough to have reached the scaling limit for these temperatures.  
To determine which one is the 
case, we must resort to a comparison between the simulation data 
with KT theory.  Figure~1(b) shows an expanded view of 
Fig.~1(a) for a few 
temperatures $23<T<30$, but for each $T$ the curve has been shifted vertically 
to remove the offset $\sigma_0^2$ so that they all coincide at $L=64$.  
The heavy dashed line indicates the KT 
slope at $T_R$ according to Eqn.(5).  The data therefore suggest that 
$T_R$ is slightly larger than 25 but less than 26, which is consistent 
with the RG prediction for $T_R = 8\pi$ in the continuum model 
\cite{80ami585,80kno597}.  
The apparent lack of an asymptotic 
$\sigma^2$ in the data for $21<T<T_R$ implies that even for $L=1024$, these 
lattice sizes are not yet large enough to be in the scaling limit for 
those temperatures.
Finally, to compare the dynamic scaling behavior of the switching algorithm 
with Metropolis, Fig.~2 shows the relaxation time in the measurement 
of $\sigma^2$ with the lattice size $L$ slightly above $T_R$.  
Compared with the dynamic exponent $\xi \approx 2.5$ in Metropolis, 
the switching algorithm shows a markedly improved $\xi \approx 1.4$.

\begin{figure}
\includegraphics[width=0.8\columnwidth]{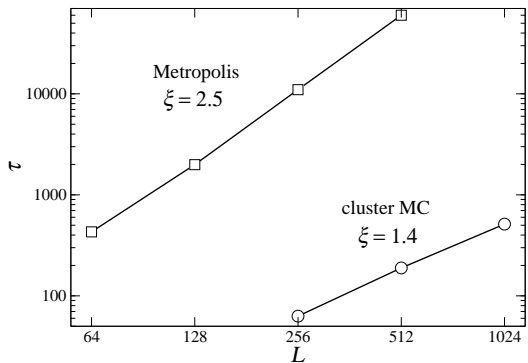}
\caption[]{
Dynamic scaling for the relaxation time $\tau$ (in units of MC passes) 
of $\sigma^2$ as a function of lattice size $L$ in Metropolis versus 
cluster MC, with their corresponding exponent $\xi$.
}
\end{figure}

\begin{acknowledgments}

This work was supported by the National Science Foundation under grant 
CHE-9970766.

\end{acknowledgments}

\end{document}